\documentclass{iau} 
\usepackage{graphicx,natbib}

\title[Quasi-Periodicities in Blazars
] 
{Quasi-Periodicities at Year Time Scales in Blazars}

\author[Stefano Covino et al.]   
{Stefano Covino$^1$,
Angela Sandrinelli$^{1,2}$
 \and Aldo Treves$^2$}

\affiliation{$^1$INAF / Brera Astronomical Observatory, via Bianchi 46, 23897, Merate (LC), Italy\\[\affilskip]
$^2$Universit\`a degli Studi dell'Insubria, Via Valleggio 11, 22100 Como, Italy}

\pubyear{2016}
\volume{324}  
\jname{New Frontiers in Black Holes Astrophysics}
\editors{A. Gombo\v{c} Editor}
\begin{document}

\maketitle

\begin{abstract}
We examine the 2008-2016 gamma-ray  and optical light curves of a number of 
bright Fermi blazars. In a fraction of them, the periodograms show 
possible evidence of quasi-periodicities related in the two bands. This coincidence strengthens their physical meaning. 
Comparing with results from the periodicity search of  quasars, the presence 
of quasi-periodicities in blazars  suggests that the basic condition for its 
observability is related to the relativistic jet in the observer direction, but 
the overall picture remains uncertain.
\end{abstract}

\firstsection 
\section{Introduction}

Blazars are active galactic nuclei with a relativistic jet pointing in the observer direction.  They are highly variable in all spectral bands \citep[e.g.][]{fal14}, and periodicities for these sorces have been hypothesized since the discovery of the first members of the class. Yet, no fully convincing case has been identified, with the possible exception of OJ\,287, with a reported 12 year period \citep[e.g.][]{Sillanpaa1988} that appears rather robust,
but not uncontroversial \citep[e.g.][]{Hudec2013}.
 
In recent years it has become also clear that blazars are the main constituents of the extragalactic $\gamma$-ray sky and the \textit{Fermi} mission since its launch has monitored the entire celestial sphere every 3 hours. Therefore, $\gamma$-ray light curves of blazars are becoming easily available \citep{Abdo2010}, opening the possibility to look for periodicities in multiple and independent bands. In addition, robotic telescopes often developed for the monitorig of high-energy transients as gamma-ray bursts, are now operational since almost a decade, providing a large amount of homogeneous data that can fruitfully be analyzed for identifying long-term oscillations.

\begin{figure}
\begin{center}
\includegraphics[width=\columnwidth]{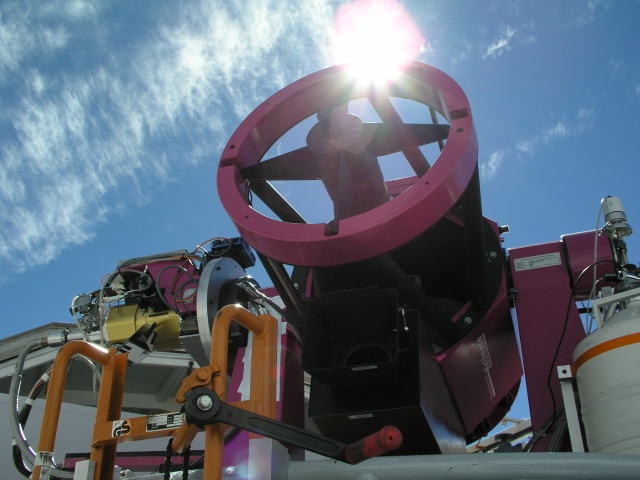}
\end{center}
\caption{The REM telescope at the ESO facility of La Silla.}
\label{figrem}
\end{figure}

\begin{figure}
\begin{center}
\includegraphics[width=\columnwidth]{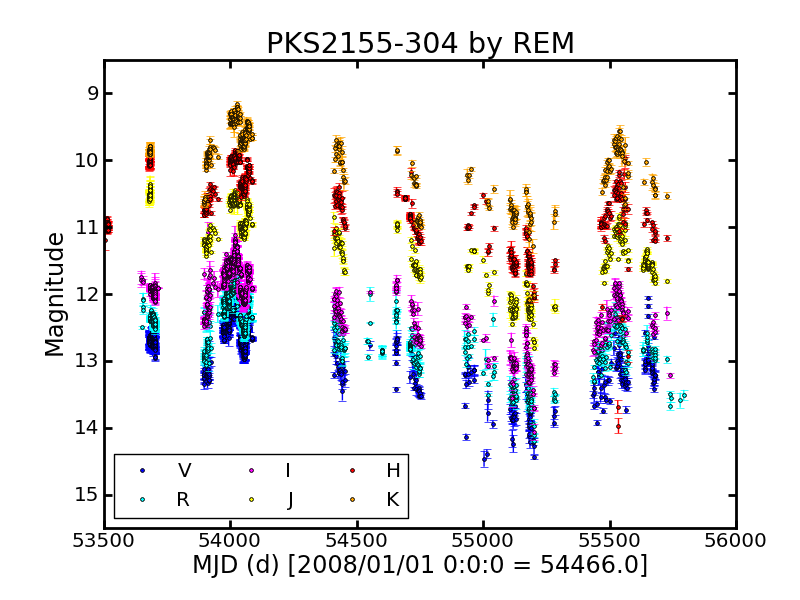}
\end{center}
\caption{Multicolor light-curve for PKS2155-304.}
\label{figlc}
\end{figure}

One of these small-size robotic facilities is the Rapid Eye Mount \citep[REM,][]{Zerbi2004, Covino2004} telescope (Fig.\,\ref{figrem}) located at the ESO premise of La Silla since 2004 and since then devoting a large fraction of the available observing time to blazar monitoring in the Southern hemisphere (Fig.\,\ref{figlc}). Our team derived and analyzed the REM data for six BL Lac objects: PKS\,0537-441, PKS\,0735+17, OJ\,287, PKS\,2005-489, PKS\,2155-304, and W\,Comae, and of the flat spectrum radio quasar PKS\,1510-089 \citep{Sandr14a}. 

\section{Quasi Periodicities}

\begin{figure}
\begin{center}
\includegraphics[width=0.9\columnwidth]{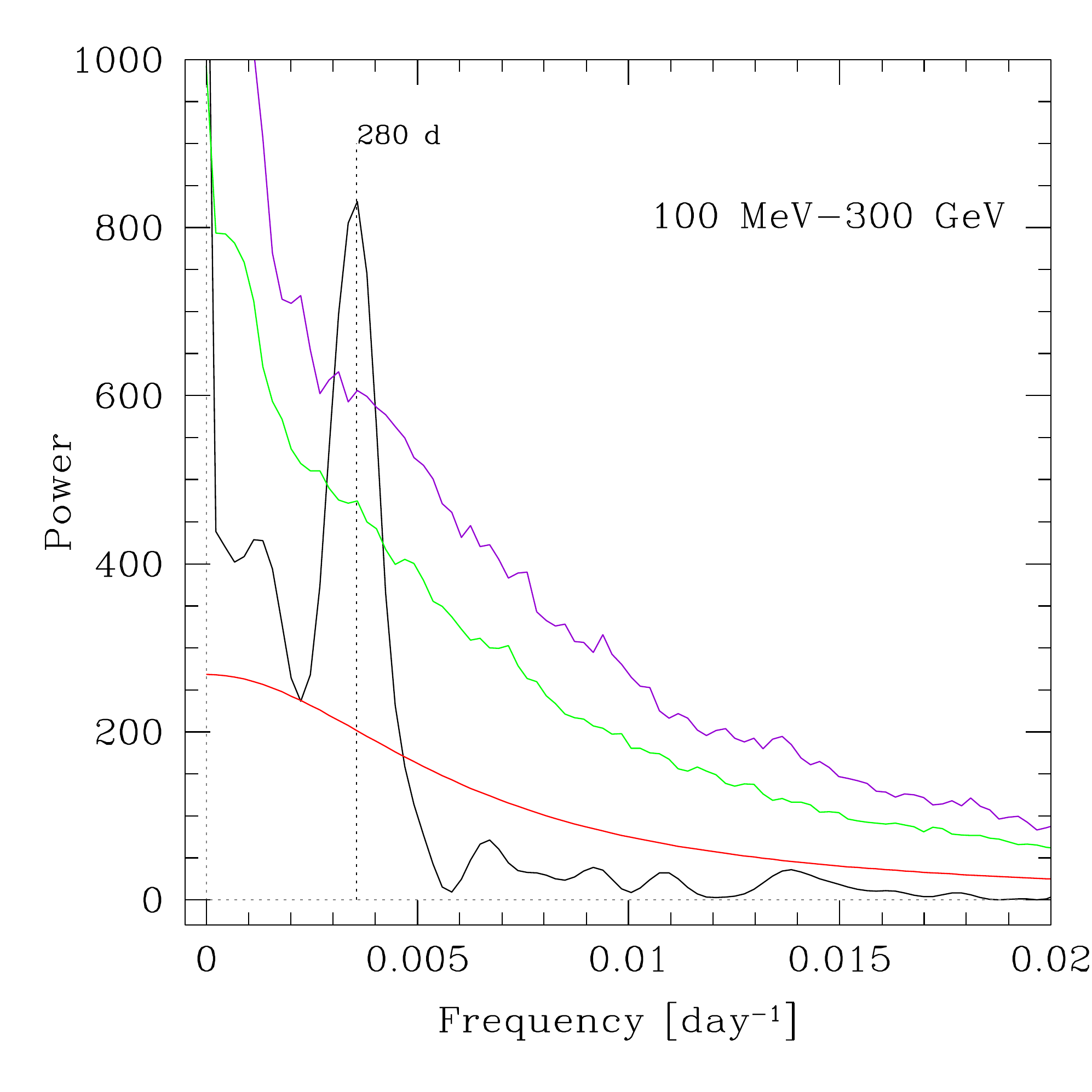}
\end{center}
\caption{Power density spectrum for the \textit{Fermi} gamma-ray data of an active phase of PKS\,0537-441 \citep[black line, from ][]{Sandr16b}. A possible periodicity at $T \sim 280$\,days is indicated. 95\% and 99\% significance curves derived from the local noise model (dashed line) are also shown.}
\label{figper}
\end{figure}

For some of the monitored sources, a search for periodicity has been carried out with optical, from REM and other facilities, and high-energy \textit{Fermi} data \citep{Sandr14b,Sandr16a,Sandr16b}. We recently also performed  analogous analyses on three BL Lac objects in the Northen sky \citep[0716+714, MRK\,421 and BL\,Lac,][]{Sandr17} and similar investigations have been proposed by other teams \citep[e.g.][]{Ack15}. Generally speaking, in a fair fraction of the analyzed sources, possible periodicities at year time-scale have been singled out (Fig.\,\ref{figper}). 

The proposed quasi-periodicities correspond to peaks in the power density spectra that stand out of the local structured noise rather clearly. However, once a correction for the ``trial-factors'' due to the large number of frequencies analyzed is applied, the global significance of the periods turns out to be rather low. The whole subject is rather complex and widely discussed in the literature \citep[e.g.][]{Horne1986,Frescura2008,Suveges14,Suveges15,Vaughan2005,Vaughan2010}, and a proper treatment of the statistical significance of the discussed quasi-periodicities will be subject of a forthcoming paper. At this stage it is enough to consider that, beyond any correct evaluation of the chance probabilities for these findings, we observe that in several cases there is a coincidence, within the errors, of the same periods in independent bands as the optical and gamma-rays. This suggests that there could be a physical meaning associated to these periods worth discussing in some detail.

\section{Discussion}

For a few sources low-significance periodicities simultaneously present in the optical and in the gamma-rays have been identified, enforcing their possible physical relevance.  The convincing cases are PKS\,2155-304 \citep{Sandr14b}, PG\,1553+11 \citep{Ack15}, PKS\,0537-447 \citep{Sandr16b} and BL\,Lac \citep{Sandr17}. These oscillations could be directly related to a real periodicity, as the orbital one of a system of two binary supermassive black holes, or indirectly as being due to an accretion disc/jet precession.  

Blazar emissions are characterized by a chaotic component with superimposed 
 a possible periodicity, which is therefore possibly hidden and difficult to single out if, as it is typically the case, the data are unevenly sampled and the time-series have of course a finite length. Furthermore, the fact that optical observation campaigns on a
 blazar are basically activated in case of a high state of the source 
introduces another complicating factor.

Recently, year-like periodicities have been searched for quasars starting
from various surveys, detecting several positive candidate cases among hundred thousands objects \citep[][and references therein]{Charisi2016}. Therefore, there could be a remarkable difference with respect to bright \textit{Fermi} blazars in terms of fraction of quasi-periodicity detections with roughly a year time-scale. If this will be proved to be true it could be related  with the main characteristic that distinguishes them from quasars, the presence of a relativistic jet  pointing in the observer direction.  The net effect would be to obviously amplify the visibility of the oscillations, since it can magnify the variability through known relativistic effects.  Furthermore, as already mentioned, it is possible that the oscillations are due to jet instabilities, independent of the presence of a binary companion.

\end{document}